\documentclass[aps,prl,twocolumn,superscriptaddress,reprint]{revtex4-1}
\usepackage{graphicx}
\usepackage{amssymb, amsmath}
\usepackage{booktabs}
\usepackage[usenames]{color}
\usepackage{bm}
\usepackage{multirow}
\usepackage{dcolumn}
\usepackage{hyperref}
\usepackage{enumerate}
\hypersetup{
    colorlinks=true,
    linkcolor=blue,
    filecolor=gray,      
    urlcolor=blue,
    citecolor=blue,
}

\begin{document}

\title{Deep-learning density functional perturbation theory}

\affiliation{State Key Laboratory of Low Dimensional Quantum Physics and Department of Physics, Tsinghua University, Beijing, 100084, China}
\affiliation{Institute for Advanced Study, Tsinghua University, Beijing 100084, China}
\affiliation{School of Physics, Peking University, Beijing 100871, China}
\affiliation{Frontier Science Center for Quantum Information, Beijing, China}
\affiliation{RIKEN Center for Emergent Matter Science (CEMS), Wako, Saitama 351-0198, Japan}

\author{He \surname{Li}}
\thanks{These authors contributed equally to this work.}
\affiliation{State Key Laboratory of Low Dimensional Quantum Physics and Department of Physics, Tsinghua University, Beijing, 100084, China}
\affiliation{Institute for Advanced Study, Tsinghua University, Beijing 100084, China}

\author{Zechen \surname{Tang}}
\thanks{These authors contributed equally to this work.}
\affiliation{State Key Laboratory of Low Dimensional Quantum Physics and Department of Physics, Tsinghua University, Beijing, 100084, China}

\author{Jingheng \surname{Fu}}
\affiliation{State Key Laboratory of Low Dimensional Quantum Physics and Department of Physics, Tsinghua University, Beijing, 100084, China}

\author{Wen-Han \surname{Dong}}
\affiliation{State Key Laboratory of Low Dimensional Quantum Physics and Department of Physics, Tsinghua University, Beijing, 100084, China}

\author{Nianlong \surname{Zou}}
\affiliation{State Key Laboratory of Low Dimensional Quantum Physics and Department of Physics, Tsinghua University, Beijing, 100084, China}

\author{Xiaoxun \surname{Gong}}
\affiliation{State Key Laboratory of Low Dimensional Quantum Physics and Department of Physics, Tsinghua University, Beijing, 100084, China}
\affiliation{School of Physics, Peking University, Beijing 100871, China}

\author{Wenhui \surname{Duan}}
\email{duanw@tsinghua.edu.cn}
\affiliation{State Key Laboratory of Low Dimensional Quantum Physics and Department of Physics, Tsinghua University, Beijing, 100084, China}
\affiliation{Institute for Advanced Study, Tsinghua University, Beijing 100084, China}
\affiliation{Frontier Science Center for Quantum Information, Beijing, China}

\author{Yong \surname{Xu}}
\email{yongxu@mail.tsinghua.edu.cn}
\affiliation{State Key Laboratory of Low Dimensional Quantum Physics and Department of Physics, Tsinghua University, Beijing, 100084, China}
\affiliation{Frontier Science Center for Quantum Information, Beijing, China}
\affiliation{RIKEN Center for Emergent Matter Science (CEMS), Wako, Saitama 351-0198, Japan}

\begin{abstract}
Calculating perturbation response properties of materials from first principles provides a vital link between theory and experiment, but is bottlenecked by the high computational cost. Here a general framework is proposed to perform density functional perturbation theory (DFPT) calculations by neural networks, greatly improving the computational efficiency. Automatic differentiation is applied on neural networks, facilitating accurate computation of derivatives. High efficiency and good accuracy of the approach are demonstrated by studying electron-phonon coupling and related physical quantities. This work brings deep-learning density functional theory and DFPT into a unified framework, creating opportunities for developing ab initio artificial intelligence.
\end{abstract}
\maketitle

Material discovery accelerated by artificial intelligence (AI) is an emerging interdisciplinary field that would profoundly change future research of materials science. An important task of this field is to create big data of materials containing comprehensive properties, preferably via high-throughput ab initio calculations. Density functional theory (DFT) is the most widely used ab initio method, by which material databases of ground-state properties are built. However, in experiments or device applications the electronic systems are inevitably perturbed away from the ground state. Density functional perturbation theory (DFPT) has been developed to predict perturbation response properties~\cite{Baroni2001, Baroni1987, Gonze1989}, such as phonons and electron-phonon coupling (EPC)~\cite{Giustino2017}, which play critical roles in a wide variety of physical phenomena, including Bardeen–Cooper–Schrieffer (BCS) superconductivity, ferroelectricity, electronic and thermal transport, infrared and Raman spectroscopy, and so on. Unfortunately, DFPT calculations are computationally quite expensive, hindering high-throughput materials research. For instance, ab initio studies of BCS superconductors typically consider systems with the number of atoms per primitive cell ($N$) no larger than 10-20~\cite{Lilia2022}, which limits the computational search of high-$T_c$ superconductors. In this context, methodological developments of DFPT are urgently demanded.

In DFPT, responses of the occupied-state manifold to perturbations are calculated by solving a set of coupled Sternheimer equations self-consistently~\cite{Martin2004}. Take perturbations of lattice vibrations for example, the complexity of solving the Sternheimer equations for each perturbation is of the same order as that for DFT, typically O($N^3$) in the Kohn-Sham scheme, and the number of relevant perturbations is proportional to $N$, leading to an overall scaling of O($N^4$). Moreover, a dense sampling of momentum space is needed to ensure convergence, which also significantly increases the computational overhead~\cite{Giustino2017}. Intensive research effort has been devoted to optimizing the method~\cite{Giustino2007, Gunst2016, Shang2017, Lin2017, Agapito2018, Chaput2019, Engel2020}, such as developing low-scaling algorithms to reduce the computational complexity of DFPT and applying Wannier interpolation or real-space techniques for efficient momentum-space sampling. These improvements help reduce the computational cost, whereas DFPT study of moderate-size systems remains challenging. Recently, AI has shown great potential to change the landscape of ab initio calculations, as demonstrated by deep-learning DFT calculations of atomic and electronic structures~\cite{Lorenz2004, Behler2007, Brockherde2017, Justin2017, Zhang2018, Schutt2018, Xie2018, Batzner2022, Musaelian2023, Li2022, Gong2023, Li2023, Li2022_2, Li2023_2}. The use of AI approaches to improve or even replace DFPT algorithms is promising but largely unexplored.

In this work, we propose a general framework to perform DFPT calculations by deep learning, which employs equivariant neural networks to learn the key quantity of DFPT---the induced change of Kohn-Sham potential per unit perturbation, trains neural networks with DFT data of random perturbations, and computes derivatives of physical quantities via automatic differentiation. We numerically implement the method for EPC calculations, and demonstrate the high efficiency and good accuracy of deep learning by example studies. The work not only paves the way for high-throughput DFPT calculations, but also unifies deep-learning DFT and DFPT into one framework, broadening the research scope of deep-learning ab initio calculation.

Perturbations are pervasive in the research of physics and materials. For instance, information of materials is detected by measuring their responses to experimental probes; devices are designed by controlling material properties with external fields. Thus calculating perturbation response properties of materials from first principles is of fundamental importance, for that DFPT is developed~\cite{Baroni2001}. In the language of Kohn-Sham DFT, a perturbation is a change of external potential $\Delta V_{\rm{ext}}$, caused by lattice vibrations, strains, electric or magnetic fields, etc. The responses of electronic systems are described by the induced change of effective Kohn-Sham potential $\Delta V_{\rm{KS}}$ or charge density $\Delta n$~\cite{Martin2004}. Compared to the bare $\Delta V_{\rm{ext}}$, $\Delta V_{\rm{KS}}$ incorporates additional changes of Hartree and exchange-correlation potentials related to $\Delta n$. Note that the variation of exchange correlation is neglected in the random phase approximation, which will not be used here.

This work will focus on perturbations of lattice vibrations, which are relevant to the study of phonons and EPC-related properties~\cite{Giustino2017}. In DFPT, a set of coupled Sternheimer equations involving occupied states are solved self-consistently for obtaining $\Delta V_{\rm{KS}}$ and $\Delta n$ to linear order in $\Delta V_{\rm{ext}}$~\cite{Baroni2001, Baroni1987, Gonze1989}. In contrast to the finite difference method and the standard perturbation theory, DFPT computes derivatives of physical quantities analytically without invoking supercells or unoccupied states~\cite{Martin2004}, and is thus more advantageous in accuracy and efficiency. A key quantity of DFPT is the derivative of $\Delta V_{\rm{KS}}$ with respect to perturbation, namely $\partial V_{\rm{KS}}$. One should consider $3N\times N_q$ independent periodic perturbations for crystalline materials, when $N_q$ wavevector points are sampled in the momentum space of phonons. This is the most computationally intensive part of DFPT. Once the full set of $\partial V_{\rm{KS}}$ is known, the perturbation response properties can be derived for any physical quantities in the single-particle picture.

\begin{figure}[t]
\includegraphics[width=1.0\linewidth]{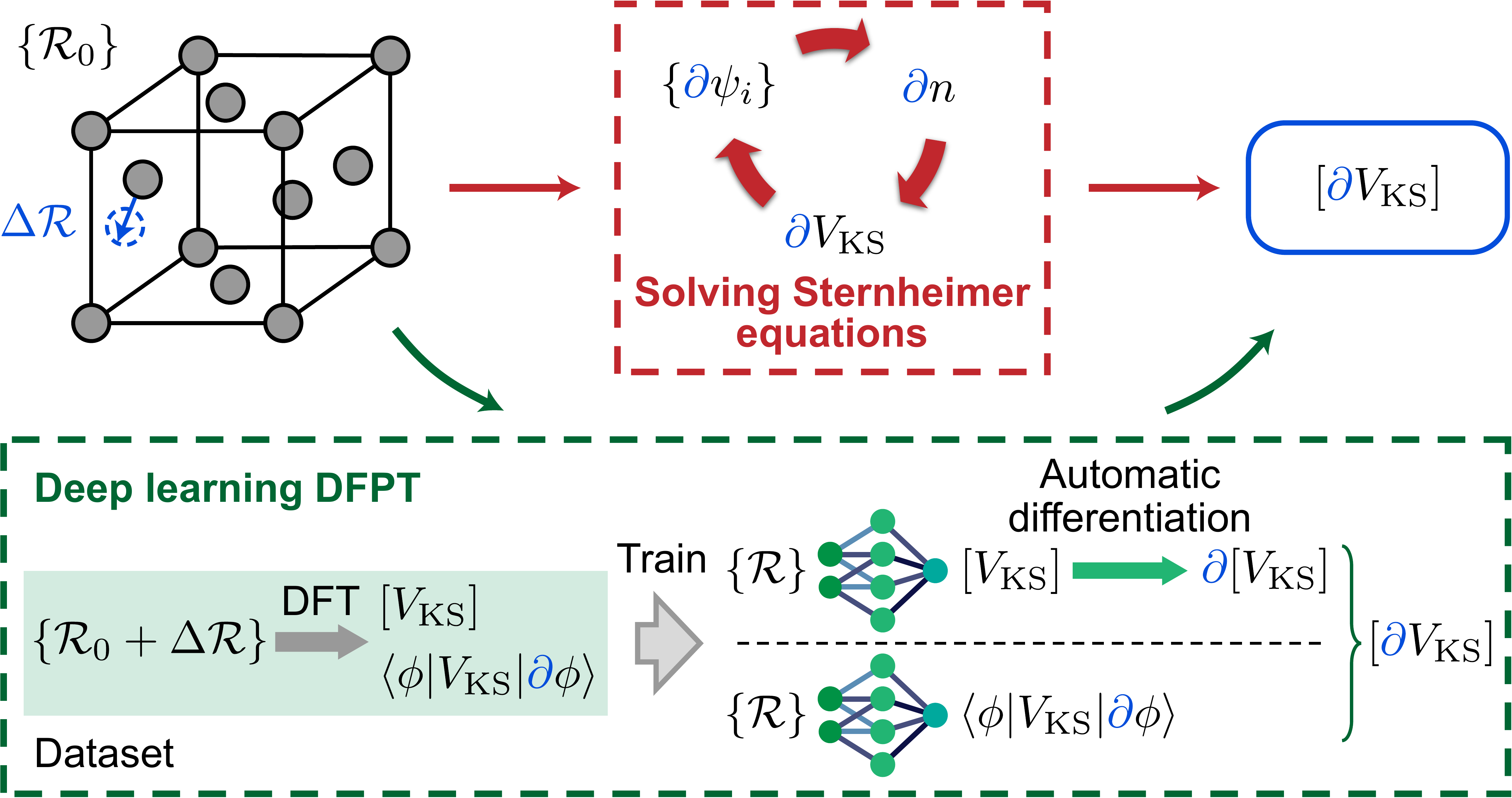}
\caption{Schematic of deep learning DFPT. The derivatives of Kohn-Sham potential, Kohn-Sham eigenstates, and charge density with respect to perturbation ($\partial V_{\mathrm{KS}}$, $\{\partial \psi_i\}$, $\partial n$) are computed by solving Sternheimer equations self-consistently in DFPT. Perturbation of atomic structure ($\Delta\mathcal{R}$) about the equilibrium configuration ($\{\mathcal{R}_0\}$) is illustrated. Using DFT data of Kohn-Sham potential matrix $[V_{\mathrm{KS}}]$ and Pulay correction terms $\langle \phi | V_{\mathrm{KS}} | \partial \phi \rangle$ for varying atomic structures, two neural network models are trained, which in combination with automatic differentiation give the electron-phonon coupling matrix $[\partial V_{\mathrm{KS}}]$.}
\label{fig1}
\end{figure}

We notice that the above problem has a special feature suitable for deep learning: A large set of $\partial V_{\rm{KS}}$ for varying perturbations are calculated about the equilibrium configuration. Such kind of perturbation information could be effectively encoded into deep neural networks, as inspired by recent studies~\cite{Li2022, Gong2023}. Importantly, derivative calculations can be efficiently and accurately on neural networks due to their differentiable nature. Considering that DFT data are more accessible than DFPT ones, we suggest to train neural-network models with DFT data of $\Delta V_{\rm{KS}}$ for random perturbations, and then perform automatic differentiation to compute the derivative quantity $\partial V_{\rm{KS}}$. By this strategy, the most time-consuming calculations of DFPT are accomplished with neural networks. This is the essential idea of deep-learning DFPT, as illustrated in Fig. \ref{fig1}.

The major task of deep learning is to represent the dependence of $\partial V_{\rm{KS}}$ on atomic structure $\{\mathcal R\}$ by neural networks. By the Hohenberg–Kohn theorem~\cite{Hohenberg1964}, $V_{\rm{KS}}$ is a function of $\{\mathcal R\}$, and so is $\partial V_{\rm{KS}}$. For solids $\partial V_{\rm{KS}}$ is often expressed in the Bloch picture: $g_{mn\nu}(\mathbf k, \mathbf q) = \langle u_{m}(\mathbf{k}+\mathbf{q})|
\partial_{\mathbf q\nu}{V}_{\text{KS}}
|u_{n}(\mathbf{k})\rangle$, where $g_{mn\nu}(\mathbf k, \mathbf q)$ is the so-called EPC matrix element, $|u_{n}(\mathbf k)\rangle$ is the initial Bloch state of the $n$-th electronic band with wavevector $\mathbf k$, $|u_{m}(\mathbf k + \mathbf q)\rangle$ is the final Bloch state of the $m$-th electronic band  with wavevector $\mathbf k + \mathbf q$, and the perturbation refers to atomic displacements induced by the $\nu$-th phonon mode with wavevector $\mathbf q$~\cite{Giustino2017}. Since the Bloch eigenstates are sensitive to distant perturbations, $g_{mn\nu}(\mathbf k, \mathbf q)$ depends on the global atomic structure, which is difficult, if not impossible, to be described by neural networks. Instead, one may employ the nearsightedness principle of electronic matter~\cite{Kohn1996,Prodan2005} to make the problem tractable. For that, we use localized atomic-like orbitals $\phi_{i\alpha}(\mathbf r)$ as basis set, where $\mathbf r$ is the coordinate choosing the $i$-th atom as the origin, $\alpha \equiv (plm)$, $p$ is the multiplicity index of radial function, $l$ and $m$ are indices of spherical harmonics $Y_{lm}$ used as angular function. This gives the real-space EPC matrix element: $g_{IJK}(\mathbf R_j, \mathbf R_k) = \langle\phi_{i\alpha, \mathbf R_0}|\frac{\partial{V}_\text{KS}}{\partial \mathcal{R}_{ka, \mathbf R_k}}|\phi_{j\beta, \mathbf R_j}\rangle$, where $I\equiv i\alpha$, $J\equiv j\beta$, and $K\equiv ka$, $\phi_{i\alpha, \mathbf R_0}$ ($\phi_{j\beta, \mathbf R_j}$) denotes the localized orbital centered at the $i$-th ( $j$-th) atom in the unit cell with lattice vector $\mathbf R_i$ ($\mathbf R_j$) using $\mathbf R_i \equiv \mathbf R_0$ as the reference, $\mathcal{R}_{ka, \mathbf R_k}$ denotes the displacement of the $k$-th atom in the unit cell with lattice vector $\mathbf R_k$ along the $a$-th ($a = x, y, z$) direction. The two kinds of EPC matrix elements are related by the formula: 
\begin{align}
\label{eq1}
\nonumber &g_{mn\nu}(\mathbf k, \mathbf q) =\sqrt{\frac{\hbar}{2M_k\omega_{\nu}(\mathbf q)}}\sum_{\mathbf R_j, \mathbf R_k}\exp\left[\mathrm i\left(\mathbf k\cdot\mathbf R_j + \mathbf q\cdot\mathbf R_k\right)\right]\\
&\times\sum_{IJK} U^*_{Im}(\mathbf k+\mathbf q) g_{IJK}(\mathbf R_j, \mathbf R_k)
U_{Jn}(\mathbf k)e_{K\nu}(\mathbf q),
\end{align}
where $M_k$ is the mass of atom $k$, $\hbar$ is the reduced Planck's constant, $\omega_{\nu}$ is the phonon frequency, $e_{K\nu}$ denotes component of phonon eigenmode, $U_{Im}$ and $U_{Jn}$ denote components of Bloch eigenstates under the localized basis. A similar formula has been derived for calculating EPC using Wannier functions~\cite{Giustino2007}.

By the nearsightedness principle, $g_{IJK}(\mathbf R_j, \mathbf R_k)$ depends on the neighboring atomic structure only, which is more suitable for deep learning than $g_{mn\nu}(\mathbf k, \mathbf q)$. In principle, one may compute $g_{IJK}(\mathbf R_j, \mathbf R_k)$ indirectly from $g_{mn\nu}(\mathbf k, \mathbf q)$ via Eq. \eqref{eq1} by the conventional DFPT, or directly by the real-space DFPT or finite-difference method. A critical problem is that $g_{IJK}(\mathbf R_j, \mathbf R_k)$ explicitly involves coupling between three sites, which has a finite but long-range cutoff in the real space, giving rise to a large amount of nonzero elements for each atomic structure. Moreover, many atomic structures will be considered in training calculations. Furthermore, even if the training data are available, training neural networks with huge amount of data is not a simple task. All these challenge the deep learning study of $g_{IJK}(\mathbf R_j, \mathbf R_k)$.

There is an elegant way to circumvent the above difficulties: Employ neural networks to learn $V_{\rm{KS}}(\{\mathcal R\})$ and then make automatic differentiation on neural networks to get $\partial V_{\rm{KS}}(\{\mathcal R\})$. Automatic differentiation is a programming paradigm widely applied in scientific computation~\cite{Ekstrom2010, Tamayo2018, Liao2019, Li2021}, which constructs programs in a fully differentiable manner and calculates derivatives of complex functions through a computation graph with machine precision. Due to the differentiable nature of neural networks, derivatives with respect to input variables of neural networks can be easily and accurately calculated by automatic differentiation. This not only simplifies the deep learning problem, but also overcomes the disadvantages of calculating derivatives numerically by finite difference or analytically by DFPT.

The real-space EPC matrix element is written as:
\begin{align}
\label{eq2}
\nonumber&g_{IJK}(\mathbf R_j, \mathbf R_k) =\frac{\partial}{\partial \mathcal{R}_{ka, \mathbf R_k}}
\langle\phi_{i\alpha, \mathbf R_0}|{V}_\text{KS}
|\phi_{j\beta, \mathbf R_j}\rangle\\
&-\langle\phi_{i\alpha, \mathbf R_0}| V_{\text{KS}}|\frac{\partial\phi_{j\beta, \mathbf R_j}}{\partial \mathcal{R}_{ka, \mathbf R_k}}\rangle
-\langle\frac{\partial\phi_{i\alpha, \mathbf R_0}}{\partial \mathcal{R}_{ka, \mathbf R_k}}| V_{\text{KS}}|\phi_{j\beta, \mathbf R_j}\rangle.
\end{align}
The first term corresponds to the derivative of Kohn-Sham potential matrix ($[V_{\mathrm{KS}}] \equiv \langle\phi|V_{\mathrm{KS}}|\phi\rangle$), namely $\partial[V_{\mathrm{KS}}]$, which will be calculated by neural networks and automatic differentiation. The second and third terms (abbreviated as $\langle\phi| V_{\text{KS}}|\partial\phi\rangle$ and $\langle \partial \phi| V_{\text{KS}}|\phi\rangle$) correspond to the Pulay corrections caused by movement of localized basis with atoms. They are complex conjugate to each other, and thus only one of them will be considered. Since $\partial\phi_{j\beta, \mathbf R_j} /\partial \mathcal{R}_{ka, \mathbf R_k}$ vanishes when $k \neq j$, the Pulay correction only involves explicit coupling between two sites, which will be computed by post processing of DFT results and used for deep learning.

\begin{figure}[ht!]
\includegraphics[width=1.0\linewidth]{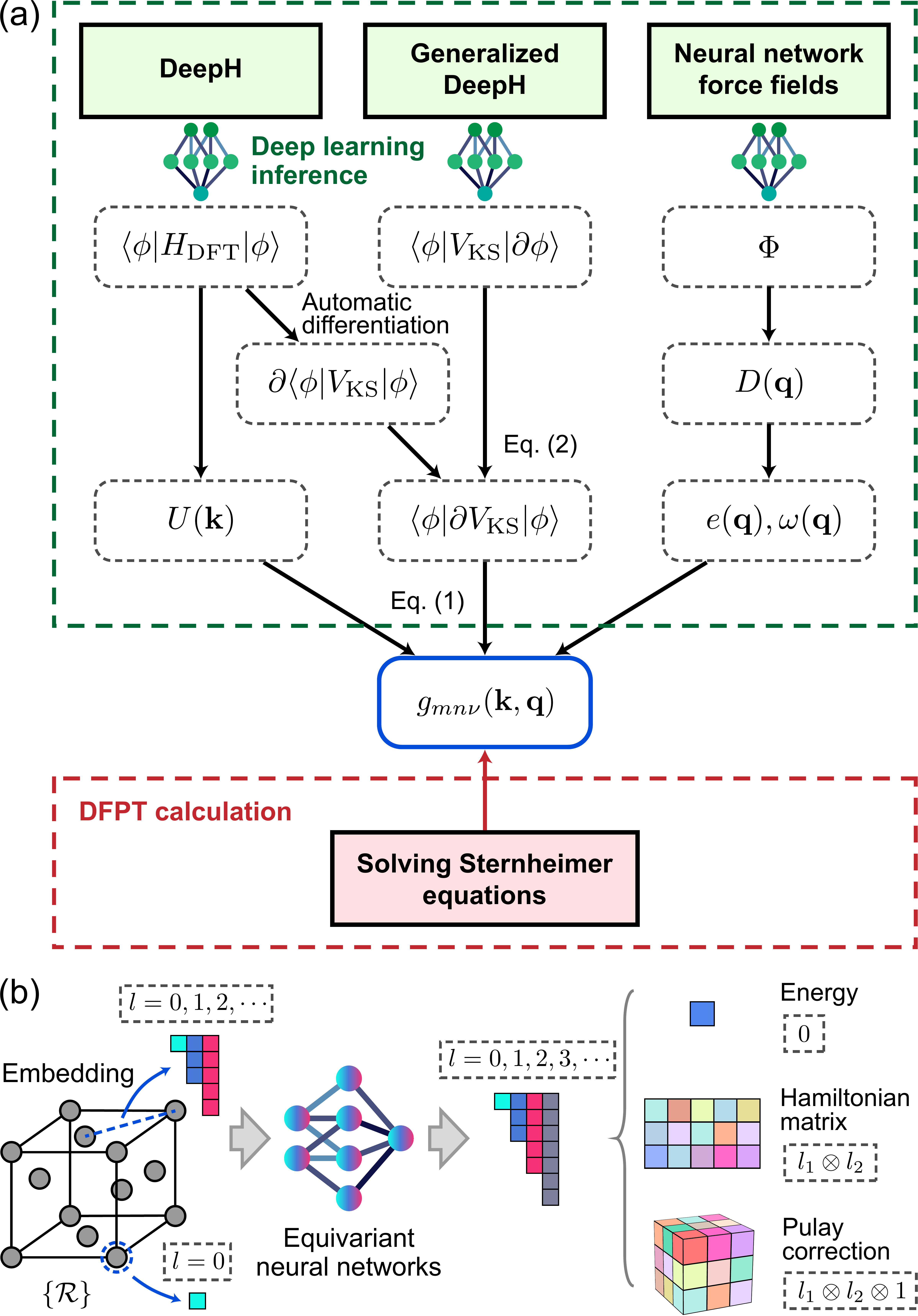}
\caption{(a) Workflow of deep learning DFPT. Three neural network models are applied to predict the DFT Hamiltonian matrix $\langle\phi|{H}_\text{DFT}|\phi\rangle$, the Pulay correction $\langle\phi| V_{\text{KS}}|\partial\phi\rangle$, and the force constant matrix $\Phi$, respectively. Automatic differentiation of the first model plus the second model gives $\langle\phi|\partial {V}_\text{KS}|\phi\rangle$. Fourier transformation of $\Phi$ yields the dynamical matrix $D(\mathbf q)$, which determines the phonon dispersion $\omega(\mathbf q)$ and eigenmode $e(\mathbf q)$. These combined with the information of Kohn-Sham eigenstates $U(\mathbf k)$ give $g_{mn\nu}(\mathbf k, \mathbf q)$, as described by Eq. \eqref{eq1}. (b) Equivariant neural networks used for representing the energy, Hamiltonian matrix, and Pulay correction terms, which are scalar, matrix, and third-order tensor, respectively. Equivariant vectors with different angular momentum quantum number $l$ (denoted by different colors) are employed to construct the matrices and tensors.}
\label{fig2}
\end{figure}

\begin{figure*}[ht!]
\includegraphics[width=0.9\linewidth]{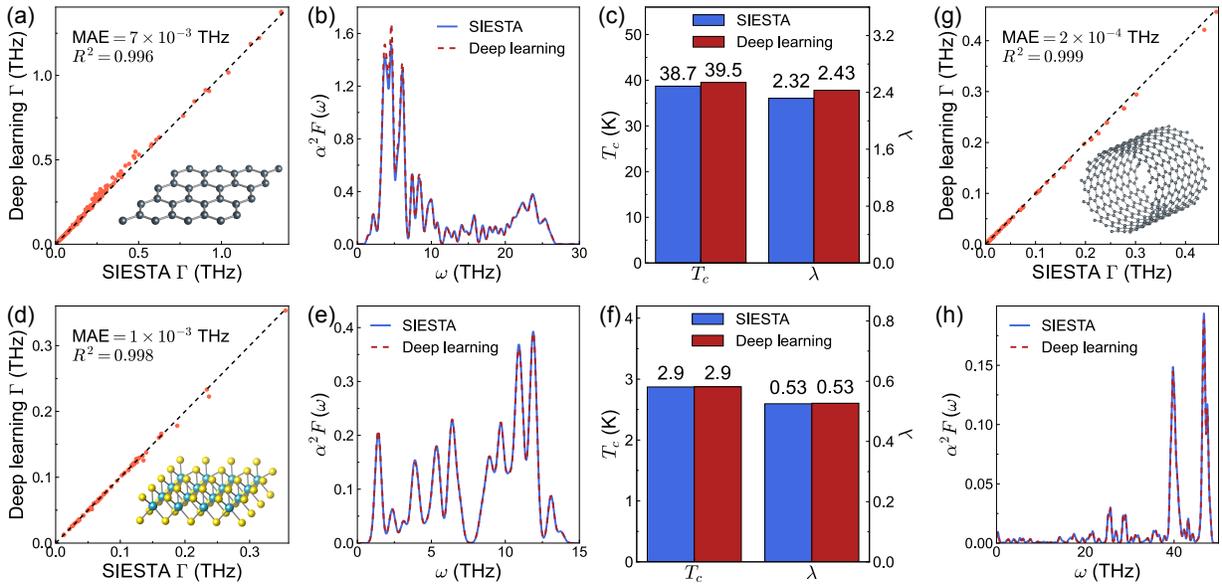}
\caption{EPC-related properties of hole-doped materials, including (a, b, c) monolayer graphene under tensile strain, (d, e, f) monolayer MoS$_2$, and (g, h) (12, 12) carbon nanotube, whose hole doping concentrations are $4.65 \times 10^{14}$ cm$^{-2}$, $4.0 \times 10^{14}$ cm$^{-2}$, and $2.3 \times 10^{6}$ cm$^{-1}$, respectively. The phonon linewidth $\Gamma$, Eliashberg spectra $\alpha^2F$ as a function of frequency $\omega$, BCS superconducting transition temperature $T_c$, and EPC strength $\lambda$ are computed by SIESTA and deep learning.}
\label{fig3}
\end{figure*}

The workflow of deep learning DFPT is shown in Fig. \ref{fig2}. Firstly, the DFT Hamiltonian matrix under localized basis $\langle\phi|{H}_\text{DFT}|\phi\rangle$ is learned by neural networks using the DeepH approach~\cite{Li2022, Gong2023, Li2023}. The generalized eigenvalue problem $H_\mathrm{DFT}U = \epsilon SU$ determines the Kohn-Sham band structure $\epsilon(\mathbf k)$ and eigenstates $U(\mathbf k)$. Here the overlap matrix $S$ and the kinetic energy matrix $\langle\phi|T|\phi\rangle$ are calculated efficiently by two-center integrals~\cite{Soler2002}. $[V_{\mathrm{KS}}]$ is obtained from $\langle\phi|{H}_\text{DFT} - T|\phi\rangle$, and then automatic differentiation is applied to get $\partial[V_{\mathrm{KS}}]$. Secondly, neural networks are applied to learn the dependence of $\langle\phi| V_{\text{KS}}|\partial\phi\rangle$ on atomic structure, as to be discussed below. $g_{IJK}(\mathbf R_j, \mathbf R_k)$ is then obtained by Eq. \eqref{eq2}. Thirdly, the phonon dispersion and eigenmodes can be calculated by using force constants predicted via neural network force fields~\cite{Unke2021}. Finally, the EPC matrix element $g_{mn\nu}(\mathbf k, \mathbf q)$ is determined by Eq. \eqref{eq1}. Thus the deep learning of DFPT is completed. Technique details are described in the Methods~\cite{supp}.

Incorporating of a priori knowledge into deep learning is of critical significance to the design of neural networks~\cite{Goodfellow2016}. Important prior knowledge for deep learning ab initio methods includes the nearsightedness principle and the symmetry principle~\cite{Li2022, Gong2023}. The nearsightedness principle has been taken into account above. The symmetry principle requires that the physical quantities and equations are equivariant under transformations of coordinate or basis, so that the laws of physics are the same for different frames of reference. This helps improve the training efficiency and prediction accuracy of neural networks. A relevant symmetry group is the Euclidean group in three dimensional space, called the E(3) group, which includes translations, rotations, and inversion. To satisfy the fundamental symmetry requirements, equivariant neural networks (ENNs)~\cite{Cohen2016, Kondor2018, Thomas2018} have been designed for the studies of DFT energy~\cite{Batzner2022, Musaelian2023} and Hamiltonian matrix~\cite{Li2022, Gong2023}. The ENN representation of Pulay correction term is rarely discussed before. In the language of group theory, $\langle\phi| V_{\text{KS}}|\partial\phi\rangle$ is composed of three-order tensors $l_1\otimes l_2 \otimes 1$ (five-order tensors $l_1\otimes \frac{1}{2} \otimes l_2 \otimes \frac{1}{2}\otimes 1$) when neglecting (including) the spin degree of freedom, where $l_1$ and $l_2$ denote the angular momentum of basis orbitals. We generalize the DeepH-E3 method~\cite{Gong2023} to represent $\langle\phi| V_{\text{KS}}|\partial\phi\rangle$ by ENNs (see details in the Supplemental Material~\cite{supp}). This generalized DeepH model employs equivariant vectors as feature vectors of neural networks, and uses the output vectors of varying angluar momentum to represent high-order tensors via the Wigner–Eckart theorem, as illustrated in Fig. \ref{fig2}(b). Up to here, a general framework of deep learning DFPT has been established.

Next we test the reliability of deep learning method by calculating EPC-related properties of three model material systems, including hole-doped monolayer graphene under tensile strain, hole-doped monolayer MoS$_2$, and hole-doped carbon nanotube. The graphene system was predicted to show high-$T_c$ BCS superconductivity~\cite{Si2013}. The DeepH and generalized DeepH models can be trained with high accuracy, showing low mean average errors (MAEs) of 0.21 meV and 0.46 meV/\AA\ (0.25 meV and 0.34 meV/\AA) in the study of graphene (MoS$_2$), respectively. Accurate prediction of EPC-related properties thus becomes feasible. As shown in Fig. \ref{fig3}(a-f), the results match well with the ab initio benchmark data. Moreover, neural network models are trained by DFT data of graphene and applied to study carbon nanotube. The latter has a curved geometry not contained in the training dataset, useful for testing the generalization ability of method. As shown in Fig. \ref{fig3}(g,h), the predicted phonon linewidth and Eliashberg spectral function of hole-doped carbon nanotube match with the benchmark results. The EPC strength $\lambda$ is predicted to be 0.085 (0.089) by deep learning (benchmark) calculations. All these systematically demonstrate good reliability of neural network methods.

\begin{figure}[t]
\includegraphics[width=1.0\linewidth]{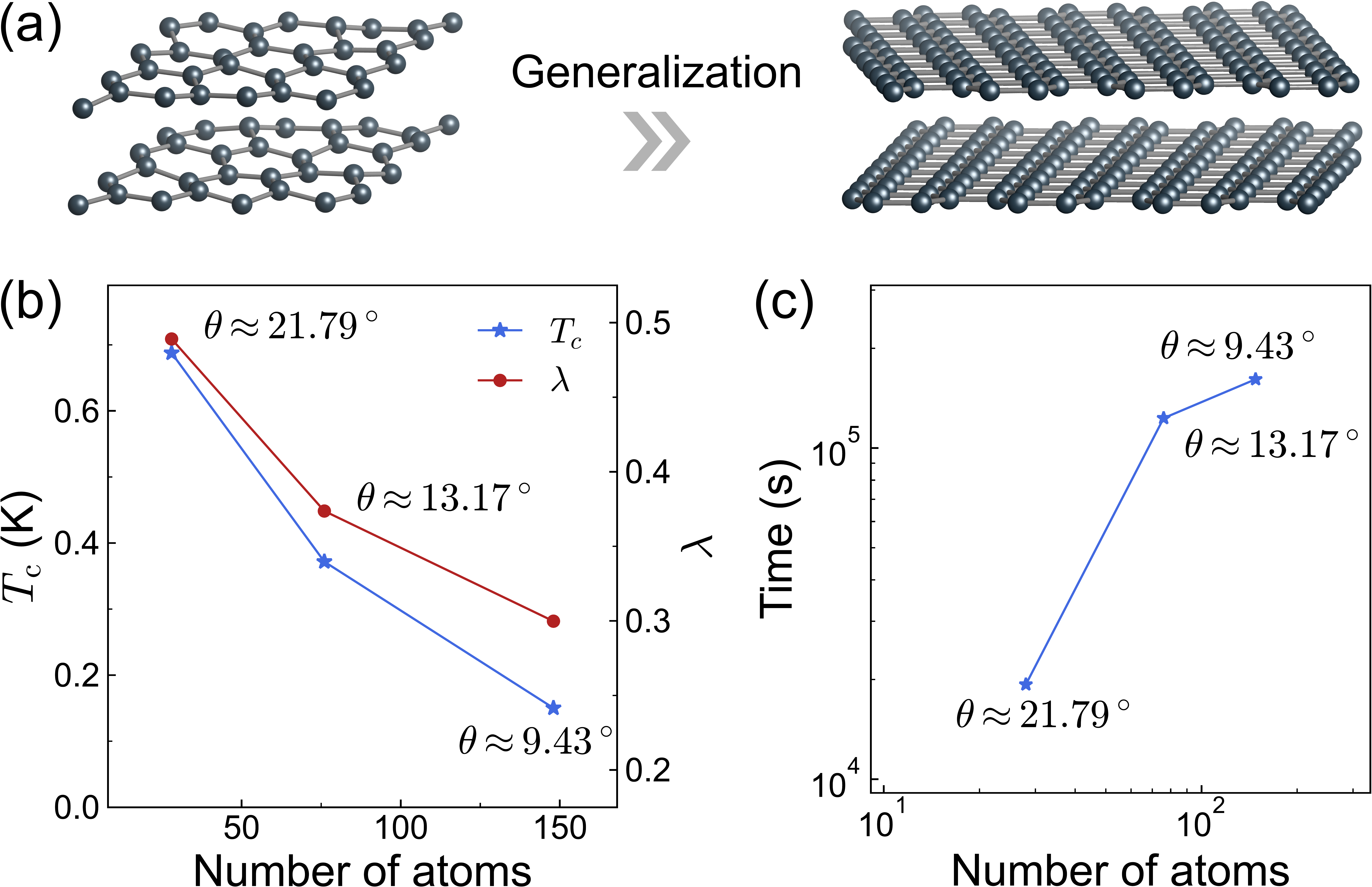}
\caption{(a) Generalization of neural networks for studying hole-doped TBGs with varying twist angles $\theta$. A hole concentration of $4.65 \times 10^{14}$ cm$^{-2}$ is considered. (b) Apparent BCS superconducting transition temperature $T_c$ and EPC strength $\lambda$ as a function of system size. (c) Computation time used for predicting the real-space EPC matrix element.}
\label{fig4}
\end{figure}

As an example application,  we apply the deep learning method to study hole-doped twisted bilayer graphene (TBG) systems (Fig. \ref{fig4}), which have attracted great research interest recently~\cite{Andrei2020}. Neural networks are trained by supercells of untwisted bilayer graphene, giving MAEs of 0.13 meV and 0.34 meV/\AA\ for the DeepH and generalized DeepH models, respectively. Moreover, neural network force fields are also trained, getting MAEs of 0.31\,meV/atom, 6.1\,meV/\AA, and 2.4\,meV/\AA$^3$, for energy, force and stress, respectively. Then material properties of TBGs with varying twist angles $\theta$, including the electronic band structure (Fig. S4), phonon dispersion (Fig. S5), EPC-related properties [Fig. \ref{fig4} (b)], can be predicted without invoking ab initio codes but fully by neural networks. Remarkably, systems containing hundreds of atoms per primitive cell are beyond the capability of conventional DFPT algorithms, but can be handled here at relatively low computational cost [Fig. \ref{fig4} (c)]. A detailed discussion of computational cost, accuracy and workflow is included in the Supplemental Material~\cite{supp}.

In summary, we develop a general framework of deep-learning DFPT, which is able to improve the computational efficiency orders of magnitude without affecting accuracy, enabling DFPT computation of large-size systems and facilitating high-throughput material calculations. The deep-learning approach not only works for studying perturbations of lattice vibrations, but also can be generalized to investigate other kinds of perturbations, such as strains and external fields. Moreover, by combining neural networks with differentiable programming, we unify deep-learning DFT and DFPT into a coherent framework. This is very helpful for future method developments, because the research of deep-learning DFT and DFPT could benefit from each other. For instance, methods have been recently developed in the framework of deep-learning DFT to deal with spin-orbit coupling~\cite{Gong2023}, magnetic materials~\cite{Li2023}, and advanced hybrid functionals~\cite{Tang2023}. Generalizing these methods to DFPT is expected to be straightforward. Furthermore, the automatic differentiation of neural networks allows efficient and accurate computation of high-order derivatives of physical quantities, which are essential to investigating various kinds of physical phenomena (like piezoelectric effects, nonlinear dielectric susceptibility, and anharmonic effects)~\cite{Giustino2017} but cannot be easily calculated by conventional methods. It is also natural to generalize the deep-learning DFPT approach for studying electric field perturbations to get a more accurate description of van der Waals interaction~\cite{Nguyen2009}. Overall, the work could significantly expand the research scope of DFPT and open new opportunities for developing deep-learning ab initio methods.

\begin{acknowledgments}
This work was supported by the Basic Science Center Project of NSFC (grant no. 52388201), the National Natural Science Foundation of China (grant no. 12334003), the National Science Fund for Distinguished Young Scholars (grant no. 12025405), the Ministry of Science and Technology of China (grant no. 2023YFA1406400), the Beijing Advanced Innovation Center for Future Chip (ICFC), and the Beijing Advanced Innovation Center for Materials Genome Engineering. The calculations were done on Hefei advanced computing center.
\end{acknowledgments}

\bibliography{reference}
\bibliographystyle{apsrev4-2}

\end{document}